# Feasibility Layer Aided Machine Learning Approach for Day-Ahead Operations

Arun Venkatesh Ramesh, *Student Member, IEEE* and Xingpeng Li, *Senior Member, IEEE*

*Abstract*— Day-ahead operation involves a complex and computationally intensive optimization process to determine the generator commitment schedule and dispatch. The optimization process is a mixed-integer linear program (MILP) also known as security-constrained unit commitment (SCUC). Independent system operators (ISOs) run SCUC daily and require state-of-the-art algorithms to speed up the process. Existing patterns in historical information can be leveraged for model reduction of SCUC, which can provide significant time savings. In this paper, machine learning (ML) based classification approaches, namely logistic regression, neural networks, random forest and K-nearest neighbor, were studied for model reduction of SCUC. The ML was then aided with a feasibility layer (FL) and post-process technique to ensure high quality solutions. The proposed approach is validated on several test systems namely, IEEE 24-Bus system, IEEE-73 Bus system, IEEE 118-Bus system, 500-Bus system, and Polish 2383-Bus system. Moreover, model reduction of a stochastic SCUC (SSCUC) was demonstrated utilizing a modified IEEE 24-Bus system with renewable generation. Simulation results demonstrate a high training accuracy to identify commitment schedule while FL and post-process ensure ML predictions do not lead to infeasible solutions with minimal loss in solution quality.

*Index Terms*— Constraint reduction, Model reduction, Variable reduction, Logistic regression, Neural network, Random forest, K-nearest neighbor, Machine learning, Mixed-integer linear programming, Security-constrained unit commitment.

NOMENCLATURE

Sets:
- $G$ — Set of generators.
- $g(n)$ — Set of generators connecting bus $n$.
- $K$ — Set of all transmission elements.
- $N$ — Set of all buses.
- $T$ — Set of Time intervals.
- $S$ — Set of Scenarios.
- $W$ — Set of renewable units.
- $w(n)$ — Set of generators connecting bus $n$.
- $\delta^+(n)$ — Set of lines with bus $n$ as receiving bus.
- $\delta^-(n)$ — Set of lines with bus $n$ as sending bus.
- $M$ — Total samples.
- $M^{test}$ — Test Samples.
- $M^{train}$ — Training Samples.

Parameters:
- $b_k$ — Susceptance of line $k$.
- $c_g$ — Linear cost for generator $g$.
- $c_g^{NL}$ — No-load cost for generator $g$.
- $c_g^{SU}$ — Start-up cost for generator $g$.
- $d_{g,t}^m$ — Predicted demand of bus $n$ in time period $t$ for generated sample m.
- $P_g^{max}$ — Maximum capacity of generator $g$.
- $P_g^{min}$ — Minimum capacity of generator $g$.
- $P_k^{max}$ — Long-term thermal line limit for line $k$.
- $R_g^{10}$ — 10-minute outage ramping limit of generator $g$.
- $R_g^{hr}$ — Regular hourly ramping limit of generator $g$.
- $R_g^{SD}$ — Shut-down ramping limit of generator $g$.
- $R_g^{SU}$ — Start-up ramping limit of generator $g$.
- $U_{m,g,t}^{ML}$ — Machine learning predicted commitment status for generator $g$ in period $t$ for sample $m$.
- $\pi_s$ — Probability of renewable energy scenario $s$.
- $P_{w,t,s}$ — Renewable output for unit $w$ in time period $t$ and scenario $s$.

Variables:
- $P_{g,t,s}$ — Output of generator $g$ in time period $t$ and scenario $s$.
- $P_{k,t,s}$ — Lineflow of line $k$ in time period $t$ and scenario $s$.
- $r_{g,t,s}$ — Reserve of generator $g$ in time period $t$ and scenario $s$.
- $u_{g,t}^m$ — Commitment status of generator $g$ in time period $t$ for generated sample m.
- $v_{g,t}$ — Start-up variable of generator $g$ in time period $t$.
- $\theta_{i,t,s}$ — Phase angle of bus $i$ in time period $t$ and scenario $s$.
- $\theta_{o,t,s}$ — Phase angle of bus $o$ in time period $t$ and scenario $s$.
- $u_{m,g,t}^{MF}$ — Feasibility layer processed commitment status for generator $g$ in period $t$ for sample $m$.
- $v_{m,g,t}^{MF}$ — Feasibility layer processed start-up status for generator $g$ in period $t$ for sample $m$.
- $u_{g,t}^{up}$ — Commitment status turned ON from OFF by feasibility layer for generator $g$ in period $t$.
- $u_{g,t}^{dn}$ — Commitment status turned OFF from ON by feasibility layer for generator $g$ in period $t$.

## I. INTRODUCTION

The short-term power system operation is a complex process which begins with day-ahead markets where generator schedules are identified for a least operational cost. Here, unit commitment is an optimization problem utilized to meet the supply and demand for tomorrow's need. The day-ahead market is responsible to schedule and commit majority of the demand requirement making it a vital step in power system operations. Since the optimization problem involves the ON/OFF status of generators, it involves binary variables and constraints making the problem a mixed-integer linear program (MILP). However, MILP makes the problem harder to solve typically for larger systems. Moreover, there are several security constraints and physical constraints to adhere with to ensure reliable and low-cost solutions. Thus, the resulting MILP is a security-constrained unit commitment (SCUC) [1]-[5]. In the deregulated regions in the United States, SCUC is solved by independent system operators (ISOs). ISOs have strict timelines to produce results for example, California ISO closes the input bids by 10:00 am and posts the schedules by 01:00 pm whereas New York ISO collects the bids by 05:00 am and posts solution by 08:00 am. This implies that the day-ahead market is cleared and the commitment schedules are provided in 3 hours [6]-[7]. Here, state-of-the-art algorithms are required to provide significant time saving benefits without loss in solution quality. Therefore, several heuristic or decomposition based algorithms were proposed to obtain the solution faster [8]-[10]. However, techniques involving machine learning (ML) to enhance SCUC were seldom studied. In comparison, learning historical information can be beneficial in reducing the complexity of the SCUC. Not only that, learning-based methods can also be used in tandem with other heuristic or decomposition to obtain further improvements.

An important factor for ML methods is the availability of good data and the right models for training to provide high quality outputs. Since the SCUC is run daily, the historical information can be leveraged to learn non-linear relationships between inputs and outputs. ML has been successfully utilized in the prediction or decision support in complex problems in various power system fields [11]-[15]. The advantage of ML is that once the model is trained the outputs can be obtained instantaneously for similar inputs. Since ML uses large amount of data to train, it can be robust to noisy data. Therefore, combining ML techniques with traditional algorithms such as SCUC can improve the overall performance [16]-[34].

The SCUC problem consists of parameters (known fixed values), variables (continuous and binary) and constraints (equalities and inequalities). The SCUC problem can have multiple feasible solutions but the optimal commitment and dispatch schedule leads to the lowest-cost solution. ML techniques and data-driven approaches have been utilized recently in aiding or replacing the SCUC process. However, most papers predominantly focus only on replacing the MILP with ML [17]-[19] or screening redundant constraints [20]-[28]. In particular, replacing SCUC with ML techniques can definitely provide the most time-saving benefits but it can never guarantee feasibility, and/or optimality. An infeasible solution is not a practical solution since several physical constraints can be breeched and [17]-[19] did not compare the solutions with the respective MILP solutions.

The papers proposing screening of constraints mostly focus only removing redundant transmission constraints in SCUC. In [20], a good starting solution was achieved for SCUC by integrating data-driven approach along with variable categorizing to improve the computational performance of SCUC. In [21], historical data was utilized to screen transmission constraints that are non-binding in the SCUC to speed up the process. Similarly, [22] uses an offline ML tool to learn about outage schedules and identifies planned outages. In [23], the authors perform a feasibility study where they mention that ML techniques cannot guarantee optimality and hence can only be used for warm-start application. The same authors in [24] then use ML techniques to identify line outages under drastic weather conditions for stochastic SCUC to eliminate congested transmission constraints. In [25], the optimization is benefitted by replacing few active and inactive constraints line-flow constraints by cost-based inequality through ML. In [26], a two-step offline and online process is implemented where the offline process screens security constraints for SCUC whereas this further screened in real-time in the security-constrained economic dispatch (SCED). Similarly, [27] performs screening only for SCED which does not bring about much time-saving benefits whereas [28] creates artificial colorful images to utilize convolutional neural networks (CNN) in SCED to study the network constraints.

Though constraint screening relaxes the SCUC algorithm when aided by ML, they cannot offer a greater time-saving benefits than variable reduction by learning the commitment schedules as seen in [29]-[34]. This is because in constraint screening the feasibility region of the SCUC solutions remain unaltered and only redundant constraints or inactive constraints are eliminated. [29]-[31] tries to eliminate all binary variables in SCUC and perform SCED. This may work for smaller systems or eliminating temporal constraints (single period application) or if the dataset is invariable which is not practical. Hence, this does not guarantee feasibility of the SCUC problem. Only [32]-[34] performs a reduced-SCUC (R-SCUC) which were also tested on large practical systems and can be considered as the state-of-the-art methods. A few machine learning techniques are proposed in [33] to use historical information to improve the performance of SCUC to solve identical instances in the future. However, [33] uses support vector machine (SVM) and k-nearest neighbor (KNN) classification algorithms to learn commitment solutions for SCUC and yet are associated with drawbacks from infeasible problems. [34] utilizes an offline ML tool to categorize load profile into different categories with a pre-determined commitment schedule from history. However, [34] provides only a feasible solution and does not guarantee optimality or high solution quality. Also the proposed methods in [32]-[34] do not address renewable generation and only works on deterministic models. Renewable energy source (RES) is addressed in [35]-[36] albeit the proposed methods only learn the varying nature of renewables to identify a most likely scenario.

Hence, we focus on building a supervised ML model to predict the commitment status of each generator $g$ in each time interval $t$ (24-hours) for day-ahead operations which are further aided by post-processes to determine the confidence of the solution. The commitment status of one implies the generator is ON whereas zero represents the generator is OFF. Ideally, a classification model can be utilized when the targets only belong to two classes, also known as binary classification. This paper extends on the preliminary idea in [32] but has several improvements and innovations to supersede the prior work. Several ML classifications were studied in this paper, namely, KNN, random forest (RF), fully connected neural networks (NN), logistic regression (LR), and a novel multi-target logistic regression (MTLR). Among these, the LR, MTLR and NN algorithm provided the most flexibility to post-process the ML outputs while also providing very high quality solutions. The proposed method in this paper focuses on innovative partial use of ML solutions with post-processing ability to reduce variables while ensuring high solution quality along with benefits of significant solve-time

reductions for SCUC. The contributions of this paper are presented as follows:
- An NN model and an innovative multi-target logistic regression (MTLR) model are utilized to leverage historic demand profiles to predict generation commitment schedule as an offline step.
- Effective post-processing methods, utilizing the ML output to reduce the variables in SCUC model achieving model-reduction, are addressed while maintaining solution quality.
- A feasibility layer (FL) is proposed to ensure feasibility of ML solution in online optimization step.
- A bus-correlated randomized profile generation (BC-RPG) method is used to obtain data to train ML models.
- The proposed FL-aided R-SCUC (R-SCUC-FL) model can work on stochastic, deterministic, or decomposed SCUC models.

The rest of this paper is organized as follows. Section II presents the SCUC model considered in this work. Section III describes the data collection procedure while Section IV models the ML architecture and the proposed methods while the results are discussed in Section V. Finally, Section VI concludes the paper.

## II. SCUC Formulation

The operational cost of generators, which includes the production, start-up and no-load costs, is minimized in the SCUC objective (1) subject to generation and power flow physical constraints. The generation constraints are modelled in (2)-(11). Here, (2) and (3) represent the minimum and maximum generation limits respectively; (4) enforces the emergency 10-min reserve ramp requirement; (5) ensures that reserves are held at the least to handle the failure of the largest system generator; (6) and (7) models the hourly generation ramp capability. Generator start-up indication variable is defined in (8)-(10) through minimum-up and minimum-down time constraints. The generator commitment status and start-up variables are binary variables as shown in (11). The base-case physical power flow constraint is represented through (12)-(14). (12) depicts the power flow calculation; (13) represents the long-term thermal limits of transmission elements; and (14) enforces nodal power balance. Slack equation, (15), is added to define the reference phase angle in the base-case solution.

*Objective:*
$$Min \sum_g \sum_t \left(c_g^{NL} u_{g,t} + c_g^{SU} v_{g,t} + \sum_s (\pi_s c_g P_{g,t,s})\right) \quad (1)$$
s.t.:
*Generation constraints:*
$$P_g^{min} u_{g,t}^m \leq P_{g,t,s}, \forall g,t,s \quad (2)$$
$$P_{g,t} + r_{g,t,s} \leq P_g^{max} u_{g,t}^m, \forall g,t,s \quad (3)$$
$$0 \leq r_{g,t,s} \leq R_g^{10} u_{g,t}^m, \forall g,t,s \quad (4)$$
$$\sum_{q \in G} r_{q,t,s} \geq P_{g,t,s} + r_{g,t,s}, \forall g,t,s \quad (5)$$
$$P_{g,t,s} - P_{g,t-1,s} \leq R_g^{hr} u_{g,t-1}^m + R_g^{SU} v_{g,t}, \forall g,t,s \quad (6)$$
$$P_{g,t-1,s} - P_{g,t,s} \leq R_g^{hr} u_{g,t}^m + R_g^{SD}(v_{g,t} - u_{g,t}^m + u_{g,t-1}^m), \forall g,t,s \quad (7)$$
$$\sum_{q=t-UT_g+1}^{t} v_{g,q} \leq u_{g,t}^m, \forall g, t \geq UT_g \quad (8)$$
$$\sum_{q=t+1}^{t+DT_g} v_{g,q} \leq 1 - u_{g,t}^m, \forall g, t \leq T - DT_g \quad (9)$$
$$v_{g,t} \geq u_{g,t}^m - u_{g,t-1}^m, \forall g,t \quad (10)$$
$$v_{g,t}, u_{g,t}^m \in \{0,1\}, \forall g,t \quad (11)$$

*Power flow constraints:*
$$P_{k,t,s} - b_k(\theta_{i,t,s} - \theta_{o,t,s}) = 0, \forall k,t,s \quad (12)$$
$$-P_k^{max} \leq P_{k,t,s} \leq P_k^{max}, \forall k,t,s \quad (13)$$
$$\sum_{g \in g(n)} P_{g,t,s} + \sum_{k \in \delta^+(n)} P_{k,t,s} - \sum_{k \in \delta^-(n)} P_{k,t,s} = d_{n,t}^m - \sum_{w \in w(n)} P_{w,t,s}, \forall n,t,s \quad (14)$$
$$\theta_{ref,t,s} = 0 \ \forall t,s \quad (15)$$

A SCUC model with relevant base case constraints are formulated to highlight the proposed ML based approach. Based on the above constraints, the deterministic SCUC formulation is represented by (1)-(15) with one scenario, $s \in \{1\}$, $\pi_s = 1.0$ and $P_{w,t,s} = 0 \ \forall g,t,$. The stochastic SCUC (SSCUC) is formulated by (1)-(15) with $N$ renewable scenarios, $s \in \{1,2,...,N\}$ and $\pi_s = \frac{1}{N}$.

## III. Data Generation

ISO's run the SCUC daily, therefore, data related to daily load-profiles and respective cleared generator commitment and dispatch schedules are stored. This data is assumed to be the starting point for this work. To train ML models, a large amount of data is required. Hence, the SCUC model specified in Section-II is utilized to generate the data. By varying the input nodal load-profile, we can generate multiple optimal commitment and dispatch schedules for respective profiles that can be collected as historical information.

It can be noted that RES can also be modelled in this step if the system has wind/solar units. The load profile then becomes a net-load profile with multiple scenarios. RES are integrated to SCUC with a multi-scenario stochastic approach in SCUC as seen in [4]-[5] and only a single resultant commitment schedule satisfies all the scenarios.

For the test systems considered in this study, the historical information is generated by modifying the nodal load profile artificially mimicking uncertainty. Since the test systems considered do not consist of the same information, a data creation step using the proposed BC-RPG method is required. To begin, a common load profile for each test system is considered with average seasonal peak information from [37]. If seasonal information are considered then average seasonal load-profile can be utilized and different ML model can be trained and stored for each season by curating the data into seasonal buckets, if needed. Once the standard profile is chosen, multiple profiles can be generated using random variables as seen in (16) where the random variables, $\alpha^m$ and $\beta_{n,t}^m$, shift the entire system load profile up/down or the composition of the system load profile can be altered, respectively. Since demand profiles only change marginally day-to-day, the value for $\alpha^m$ is considered to be $\pm 10\%$. Nodal values cannot be altered significantly as this would lose the correlation of nodal information. Therefore $\beta_{n,t}^m$ is considered to be $\pm 4\%$. The combination of both random variables provide varying load-profile curves. From Fig. 1, for example, curve 1 represents the initial load profile whereas curve 2 and curve 3 are generated only using only $\alpha^m$, curve 4 is generated through only $\beta_{n,t}^m$, and curve 5 and curve 6 are generated using the combination of both random variables.

$$SysD_t^m = \sum_{n \in N} d_{n,t}^m = (\sum_{n \in N}(d_{n,t} + \beta_{n,t}^m d_{n,t})) * (1 + \alpha^m), \forall m \in M, t \in T \quad (16)$$

Where,
$SysD_t^m$ is the system demand in time period $t$ for sample $m$.
$\alpha^m$ is a random variable ($\pm 10\%$) for sample $m$.
$\beta_{n,t}^m$ is a random variable ($\pm 4\%$) for sample $m$ for bus $n$ in time period $t$.

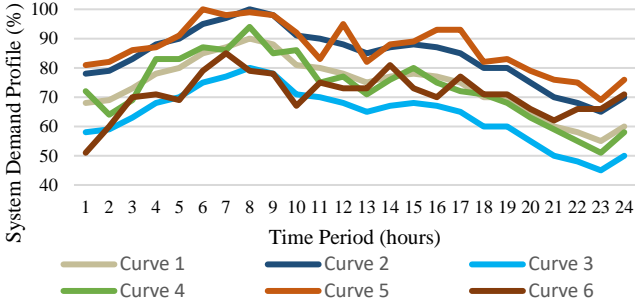

Fig. 1. Sample demand profile curves.

Only feasible samples of the SCUC are considered and a total of 1500 samples denoted as $M$ are created for each test system, which can be considered as an equivalent of 3-4 years of data. The created $M$ samples, once shuffled, are split into two datasets: 80% training samples (1200 samples) $M^{train}$ and 20% testing samples (300 samples) $M^{test}$.

## IV. PROPOSED OFFLINE APPROACH

### A. ML Approach (Model Reduction)

The ML step is utilized to reduce the number of variables in the SCUC model. The traditional approach is to utilize all the information such as constants, continuous and binary variables in an online SCUC model as shown in Fig. 2. However, we can train an ML algorithm to identify variables that follow a pattern, especially binary variables by leveraging historical information. It is known that binary variables increase the complexity in an MILP [17]-[18], [29]-[34].

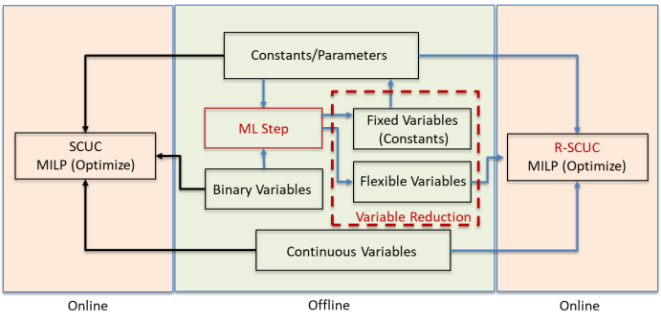

Fig. 2. SCUC Model Reduction.

In SCUC, the binary variables are the generator commitment schedule. The constants include forecasted load profiles, generator cost and ramping information whereas the continuous variables are generator dispatch, line flows and bus angles. By studying the historical commitment schedules with respective load profiles, the ML algorithm can identify many generator states with certainty for any given load profile. The generator states can be classified as either (i) flexible, to be determined by online optimization step, or (ii) fixed, as identified by offline ML algorithm. Therefore, the fixed generators are now constants and the resultant R-SCUC online model only determines the states of flexible generators.

It can be noted here that this approach of model reduction is agnostic to the MILP model. This implies that the proposed approach of model reduction is unaltered and can be applied to deterministic, stochastic and/or decomposition or heuristic techniques based SCUC models.

### B. Classification Models from Scikit-learn

In this paper, we initially compare the performances of several classification model, namely KNN, RF, NN and LR. All the models used for comparison are obtained from Scikit package, [38]. The neighbors-based classification is a type of instance-based learning or non-generalizing learning. This implies a general internal model is not constructed but rather training data are stored as instances. Classification is computed from a simple majority vote of the nearest neighbors of each point. For KNN classification, the optimal choice of the value K is highly data-dependent: in general, a larger K suppresses the effects of noise, but makes the classification boundaries less distinct. KNN works by identifying the most similar examples in the training dataset and conducting a simple majority vote [39].

Another supervised learning method used for classification is the class of non-parametric decision trees where a target variable is predicted by the model by learning simple decision rules inferred from the data features. Here, a tree can be seen as a piecewise constant approximation. The RF model is a classification where an ensemble of many individual decision trees is used for prediction. Each individual tree in the random forest predicts a class output and the class with the most votes becomes the RF prediction [40].

LR is a well-established classifier method which works in a one vs rest classification meaning it identifies one output based on the set of training inputs [32]. LR algorithm using the solver liblinear which uses a coordinate descent algorithm and only supports binary classification. The package is capable of handling one target or output at a time and the regularization is applied by default [41]. The architecture of LR is represented in Fig. 3.

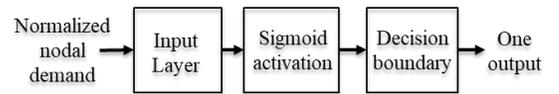

Fig. 3. LR architecture.

Multi-layer perceptron (MLP) is a supervised learning algorithm that learns a mapping between inputs and outputs by training on a dataset. MLP is also known as NN where a non-linear function approximator for either classification or regression is used for learning. Mainly, NN differs from LR, in that between the input and the output layer, there can be one or more non-linear layers hidden layers. The outputs from the hidden layer are processed by a sigmoid layer to provide probability estimates and followed by a classification to represent the class. It uses a cross-entropy loss function and trains via backpropagation. For classification, it minimizes the cross-entropy loss function, providing a vector of probability estimates [42]. The architecture of NN is shown in Fig. 4.

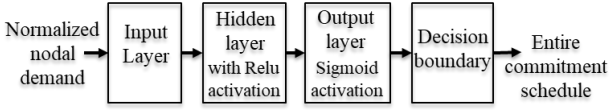

Fig. 4. NN architecture.

## C. Multi-Target Logistic Regression

In this section, the training model/algorithm considered is an MTLR model as denoted in Fig. 5. This model is similar to LR as it is a regression model which predicts the value of probability of an output being 1 [43]-[44]. The difference is that MTLR uses a single set of weights, $w_j$, as opposed to multiple models with different weights, $w_{j,m}$, in LR.

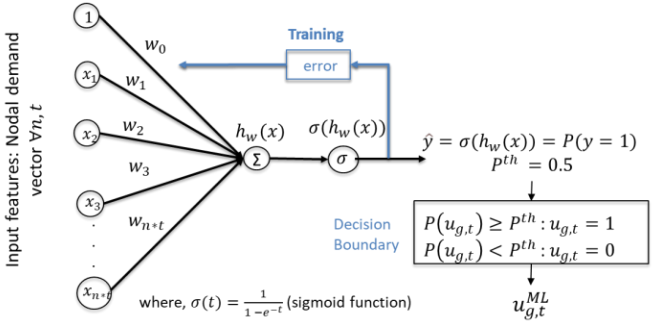

Fig. 5. MTLR based SCUC model reduction algorithm.

The hypothesis of LR model, (17), is a linear summation of normalized nodal demand and the parameters/weights, $w_j$.

$$h_w(x) = w_0 + \sum_{j \in N*T} w_j x_j \quad (17)$$

The LR model uses a sigmoid activation layer, (18), which restricts the output from 0-1 which represents the probability of the output being 1, $P(y = 1)$.

$$\sigma(t) = \frac{1}{1-e^{-t}} \quad (18)$$

Finally, the output, $\hat{y}$, is obtained after the hypothesis function followed by the sigmoid activation as seen in (19).

$$\hat{y} = \sigma(h_w(x)) = P(y = 1) \quad (19)$$

To train the LR model, we need to obtain the best parameters, $w_j$, that fit the input and output features. This is implemented using the LR cost/loss function, (20).

$$J(h_w(x)) = -\frac{1}{m}\left[\sum_{i=1}^{m}\left(y^{(i)} \log h_w(x^{(i)}) + (1-y^{(i)}) \log(1-h_w(x^{(i)}))\right)\right] \quad (20)$$

To obtain the weights, we minimize the LR cost/loss function, (21), by using a gradient descent algorithm, (22), for several iterations until the cost/loss values saturates for all samples in $M^{train}$. Here, $\delta$ represents the learning rate of the gradient descent algorithm. The number of iterations and learning rate represents the hyper-parameters of the LR model.

$$\min_w J(h_w(x)) \quad (21)$$

$$\text{Repeat } \{\omega_i := \omega_i - \delta \sum_{i=1}^{m}(h_w(x^{(i)}) - y^{(i)})x_j^{(i)}\} \quad (22)$$

The model accuracy can be verified using the post-processed outputs. Once the model is trained, the output probabilities are post-processed as $P \geq 0.5$ as 1 and $P < 0.5$ as 0 to obtain the predicted commitment schedule, $u_{i,g,t}^{ML}$. The accuracy is calculated for both $m \in M^{train}$ and $m \in M^{test}$ using (23). Here, $N$ represents the number of respective parameters.

$$Acc = 1 - \frac{1}{N_m * N_g * N_t}\sum_{i=1}^{N_m}(\sum_{g \in G}\sum_{t \in T}|u_{i,g,t} - u_{i,g,t}^{ML}|) \quad (23)$$

## D. Feasibility Layer

Once the ML model provides the classification results, a FL is added to avoid any erroneous commitment schedules in ML outputs, $U_{m,g,t}^{ML}$, by making minor but necessary corrections across time period $t \in T$ as defined in (24). Here, if $U_{m,g,t}^{ML} = 1$, then it can be turned off with $u_{g,t}^{Dn} = 1$. Similarly, if $U_{m,g,t}^{ML} = 0$, then it can be turned on with $u_{g,t}^{Up} = 1$ as shown in (25). The FL ensures that minimum up/down time limit constraints (8)-(9) are not violated by reforming them as (26)-(27). (28) defines the respective start-up variable, $v_{g,t,m}^{MF}$. Finally, (29) ensures that the flexible generator can either be turned on or turned off whereas (30) describes the variables are bound by binary integrality. The FL is represented by (24)-(30) and is solved in the online phase. Therefore, it is performed for each generator $g$ independently per sample $m \in M^{test}$ during the verification process. Here, it can be noted that Always ON/OFF as determined by ML outputs, $u_{m,g,t}^{ML}$, in each sample $m$ can be excluded as they are already feasible for minimum up/down constraints. The solve time for FL for each generator $g$ is aggregated and added to the respective R-SCUC solve time for each sample $m$.

*Objective:*

$$\text{Min } \sum_t(u_{g,t}^{Up} + u_{g,t}^{Dn}) \quad (24)$$

*s.t.:*

$$u_{m,g,t}^{MF} = U_{m,g,t}^{ML} + u_{g,t}^{Up} - u_{g,t}^{Dn} \forall t \quad (25)$$

$$\sum_{p=t+1}^{t+DT_g} v_{g,p}^{MF} \leq 1 - u_{m,g,t}^{MF} \forall t \leq nT - DT_g \quad (26)$$

$$\sum_{p=t-UT_g+1}^{t} v_{g,p}^{MF} \leq u_{m,g,t}^{MF} \forall t \geq UT_g \quad (27)$$

$$v_{m,g,t}^{MF} \geq u_{m,g,t}^{MF} - u_{m,g,t-1}^{MF} \forall t \quad (28)$$

$$u_{g,t}^{Up} + u_{g,t}^{Dn} \leq 1 \forall t \quad (29)$$

$$u_{g,t}^{Up}, u_{g,t}^{Dn}, v_{m,g,t}^{MF}, u_{m,g,t}^{MF} \in \{0,1\}, \forall t \quad (30)$$

## E. Post-Process Technique

The LR, MTLR and NN models presented in Section IV-B and Section IV-C are extended with a post-processing technique which includes the FL described in Section IV-D as seen in Fig. 6. Since the ML outputs of the above models are probabilities of generator being ON, a decision boundary of $P^{th} = 0.5$ is used to classify ON and OFF status of generators. This implies, the generator status $u_{m,g,t}^{ML} = 1$ if $P(u_{m,g,t}^{ML}) \geq P^{th}$ or $u_{m,g,t}^{ML} = 0$ if $P(u_{m,g,t}^{ML}) < P^{th}$. Since this would lead to inaccuracies along the decision boundary which in-turn lead to infeasible solutions, the outputs are further checked for feasibility using the FL, discussed in Section IV-D. The following steps are used to complete the post-process technique for each training sample $m$:

- *Step 1*: Identify always ON/OFF generators using $u_{m,g,t}^{ML}$. If a generator $g$ is always ON ($P(u_{m,g,t}^{ML}) \geq 0.95$) in each $t \in T$ then $fix\, u_{g,t}^m = 1$ for all periods for the corresponding generator. Similarly, if the generator $g$ is always OFF ($P(u_{m,g,t}^{ML}) \leq 0.05$) in $t \in T$ then $fix\, u_{g,t}^m = 0$ for all periods for the corresponding generator.

- *Step 2*: for remaining generators after Step 1, run FL. If $P(u^{ML}_{m,g,t}) \geq 0.90$ or $P(u^{ML}_{m,g,t}) \leq 0.10$ and $u^{ML}_{m,g,t} = u^{MF}_{m,g,t}$ then generator $g$ in time-period $t$ has a fixed state, $fix\ u^m_{g,t} = u^{MF}_{m,g,t}$.
- *Step 3*: If generator $g$ in time-period $t$ is identified as a flexible generator, i.e. $0.1 < (u^{ML}_{m,g,t}) < 0.9$ or if $u^{ML}_{m,g,t} \neq u^{MF}_{m,g,t}$ then warm-start $u^m_{g,t}$ with $u^{MF}_{m,g,t}$.

For each sample $m \in M^{test}$, the above steps are implemented and the respective R-SCUC is solved to verify the quality of the ML solution. The overall flow of the process is represented in Algorithm 1.

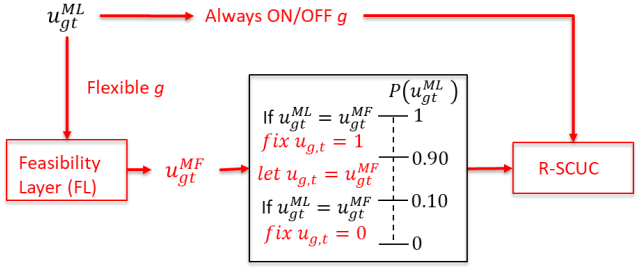

Fig. 6. Post-process technique with FL.

| **Algorithm 1** ML assisted R-SCUC-FL process |
|---|
| 1:  **Repeat** |
| 2:    randomize nodal demand |
| 3:    Solve SCUC |
| 4:    Store $d^m_{n,t}, u^m_{g,t}$, objective and time |
| 5:  **until** $m \in M$ |
| 6:  **Shuffle** $M$ samples |
| 7:  **Split** $M$ as 80% for $M^{train}$ and 20% for $M^{test}$ |
| 8:  **Train** ML using $M^{train}$ for different hyper-parameters |
| 9:  Calculate train and test accuracy |
| 10: **Tuning**: identify hyper-parameters with best accuracy |
| 11: Save ML predicted output probabilities |
| 12: **Repeat** |
| 13:   Perform *step 1-step 3* and verify R-SCUC using $u^m_{g,t}$ |
| 14:   record objective and time |
| 15: **until** $m \in M^{test}$ |

## V. RESULTS AND ANALYSIS

The mathematical model is formulated in AMPL. The data creation and verification steps are implemented using AMPL and solved using Gurobi solver. For ML step is implemented in Python 3.6. A computer with Intel® Xeon(R) W-2295 CPU @ 3.00GHz, 256 GB of RAM and NVIDIA Quadro RTX 8000, 48GB GPU was utilized.

TABLE I. TEST SYSTEM SUMMARY

| System | Gen cap (MW) | # bus | #gen | # branch |
|---|---|---|---|---|
| IEEE 24-Bus [37] | 3,393 | 24 | 33 | 38 |
| Modified IEEE 24-Bus [5] | 3,793 | 24 | 35 | 38 |
| IEEE 73-Bus [37] | 10,215 | 73 | 99 | 117 |
| IEEE 118-Bus [45] | 5,859 | 118 | 54 | 186 |
| Synthetic grid (SG) [46] | 12,189 | 500 | 90 | 597 |
| Polish [45] | 30,053 | 2,383 | 327 | 2,895 |

The following standard test systems summarized in Table I were utilized for results and analysis. It can be noted that, a modified IEEE 24-bus system was also introduced with 2 additional renewable units with a peak capacity of 200MW each. Three scenarios with varying renewable outputs are considered in the modified IEEE 24-bus system.

### A. Comparison with Scikit-learn Packages:

There are several classification techniques currently available. By utilizing scikit-learn package, we were able to compare some common classification techniques, namely, KNN, RF, LR, and a fully connected two layer neural network NN on the IEEE 24-Bus system data. The models were trained using $M^{train}$ and tested on $M^{test}$. The accuracy is calculated using (23).

TABLE II. CLASSIFICATION MODEL COMPARISON

| Classification model | Training accuracy | Testing accuracy | Infeasible test samples |
|---|---|---|---|
| LR | 97.95% | 97.55% | 43.14% |
| NN | 97.48% | 97.46% | 44.48% |
| KNN | 97.48% | 97.42% | 41.14% |
| RF | 97.31% | 97.28% | 47.16% |

From Table II, it can be noticed that all the classification methods fare well for commitment outputs. LR provides the highest accuracy followed by NN, KNN, and RF, respectively. To verify which model results in identifying more accurate sequences, we implement SCED. From Algorithm 1, SCED can be implemented by replacing *step1-step3* as $fix\ u^m_{g,t} = u^{ML}_{m,g,t}\ \forall\ m \in M^{test}, g \in G, t \in T$. By performing SCED using ML solutions, we understand that ML models cannot accurately identify the sequences and may either result in sub-optimal solutions or infeasible solutions. Therefore, ML cannot completely replace the SCUC. However, we realized that the accuracy alone is not the best metric since KNN has lower accuracy than LR and NN but results in fewest infeasible samples in comparison. RF has the lowest accuracy and this is also seen in the number of infeasible cases.

TABLE III. CONFUSION MATRIX COMPARISON

| Classification Model | True + | True − | False + | False − |
|---|---|---|---|---|
| LR | 50.47% | 47.01% | 1.25% | 1.20% |
| NN | 50.29% | 47.16% | 1.17% | 1.38% |
| KNN | 50.77% | 46.65% | 1.67% | 0.91% |
| RF | 50.23% | 47.04% | 1.28% | 1.44% |

On studying the results further, Table III summarizes the confusion matrix respective to the result in Table II. A confusion matrix provides an idea on the number of true predictions and false predictions in $M^{test}$. For any sample $m \in M^{test}$, if the predictions are accurate and entire sequence is identified implies the optimal solution is predicted. But if the number of false negatives increases, this implies that generator $g$ in period $t$ is identified as OFF instead of ON. This directly affects the number of feasible samples as the flexibility in the system in the system is lost and constraints are violated, especially (6)-(10). Not only that, as the number of false positives increase, this implies that the respective generator $g$ in period $t$ was identified as ON but in reality, it should be OFF. This affects the solution quality as sub-optimal generators or generators with insufficient capacity maybe turned ON.

However, it can also be noted that ML does provide a high number of accurate predictions in each sample. Therefore,

identifying a post-procedure may be beneficial to selectively utilize ML solutions that are known in high confidence. In order to leverage this, probability is a great metric. However, KNN and random forest models are inherently classification only model. This implies that the outcomes belong in one of several classes as generator schedules are grouped together in unique schedule buckets. Hence, these models cannot provide a probability for individual generators being ON/OFF. Models such as LR are inherently probabilistic in nature as the outputs are probability before the decision boundary is utilized to classify the outputs. Similarly, NN can also prove probabilistic outputs when a sigmoid layer is utilized. This implies that LR and NN are more flexible in nature to partially utilize the ML solution for variable reduction. This means that the probability can be construed as a true trained outcome of the ML model. Among the two models, LR results in lower false negative predictions which leads to higher accuracy and fewer infeasible cases and hence chosen as the classification model for further analysis.

### B. Comparison between LR and MTLR

Even though ML training is an offline step, while training larger models LR required high amount of training time. LR is traditionally developed as one vs rest algorithm which implies that the existing packages for LR only performs for each target (generator $g$ in time period $t$) separately [43]-[44]. This is computationally expensive since this requires training multiple sets of weights for each generator $g$ in each time period $t$ in the test systems. Hence, we proposed the MTLR described in sub-section IV.C. In comparison, the proposed MTLR provides outputs for all targets (generators) using single set of weights.

For validation of the proposed MTLR, we compared accuracy using LR from Scikit-learn package [38]. From Table IV, as the model size increases, the training time significantly increases as noted in the polish system which requires 6,178 seconds (~1.7 hours) to train. But in MTLR we notice that a minor trade-off in accuracy results in 2.4x speedup over LR on average across all test systems. This results in a significant computational speed-up during offline training in larger test systems. Not only that, LR method from scikit-learn only works if the training set contains both ON/OFF samples for each generator which implies that LR can only be applied for generators showing a flexible trend. In practicality, there can be few generators such as base plants and hydro plants that are ON irrespective of the load profile over the horizon and/or in all data samples. An assumption is required for such generators per the generic trend. In comparison, the proposed MTLR method can handle this certainty in schedules for fixed generators since true label is a vector of schedules of all generators in each sample and the entire schedule can be unique in nature. Hence, for these reasons, MTLR is used in subsequent results.

TABLE IV. VALIDATION OF MTLR ON IEEE 24-BUS SYSTEM

| # Bus | LR Test Accuracy | LR Train Time (s) | MTLR Test Accuracy | MTLR Train Time (s) |
|---|---|---|---|---|
| 24 | 97.55% | 16.19 | 97.44% | 8.18 |
| 73 | 95.39% | 374.2 | 95.96% | 176.29 |
| 118 | 95.98% | 344.85 | 95.52% | 143.21 |
| 500 | 98.87% | 743 | 98.80% | 339.63 |
| 2383 | 98.18% | 6,178.33 | 98.17% | 2,445.28 |

### C. MTLR Hyper-parameter Tuning

Each test system is trained using the MTLR model separately by utilizing the respective generated data, $M^{train}$. During training, the samples are considered as a single full batch for $m \in M^{train}$. The best trained hyper-parameters with highest accuracy will be utilized for further tests. For each test system, the hyper-parameter learning rate ($\delta$) is varied from 0.001-0.05. For each $\delta$, systems were trained until the cost saturates and then the accuracy was then calculated using (23).

During training, the cost vs iterations or epoch is registered to plot learning rate ($\delta$) graph. The $\delta$ graph represents the loss/cost with respect to the iteration which provides information about the training when different hyper-parameter is utilized. Here, Fig. 7, represents the learning curve for IEEE 24-Bus system when trained for 1000 iterations to show the saturation of the cost. For $\delta \geq 0.03$, the training cost never saturates which implies that the step is too large. For $\delta \leq 0.001$, it is slower to converge in training which implies the step is too small. The accuracy is also calculated for each $\delta$ using (23). Between $0.003 \leq \delta \leq 0.01$, the $\delta = 0.01$ is chosen for the IEEE 24-Bus system which provides the highest training accuracy and a strictly decreasing curve for learning rate.

In comparison, the scikit-learn models are trained using standard parameters provided by the package which includes an adaptive learning rate and early stopping functionality.

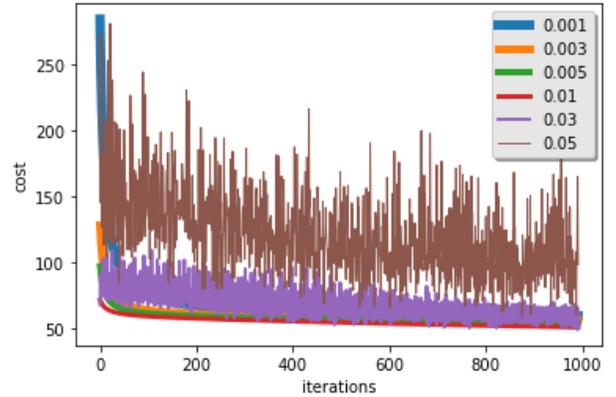

Fig. 7. Learning rate ($\delta$) curves for IEEE 24-Bus system ($0.001 \leq \delta \leq 0.05$).

### D. Training Summary (Offline)

Each system is trained using both MTLR and NN model using the respective system data, $M^{train}$, as described in section III. For all test systems, as described in Section III, 1500 samples were created, shuffled and split 80% for training and 20% for testing. The training is an offline step performed once for each system.

TABLE V. MTLR MODEL TRAINING SUMMARY

| # Bus | MTLR Accuracy | | NN Accuracy | |
|---|---|---|---|---|
| | Train | Test | Train | Test |
| 24 | 97.50% | 97.44% | 97.48% | 97.46% |
| 73 | 95.97% | 95.96% | 95.37% | 95.30% |
| 118 | 97.57% | 95.52% | 97.83% | 97.62% |
| 500 | 98.81% | 98.80% | 99.06% | 99.04% |
| 2383 | 98.34% | 98.17% | 98.11% | 97.98% |

During training, the samples are considered as a single full batch. For MTLR, the hyper-parameter learning rate ($\delta$) is identified and trained as per sub-section V.D. The training and testing accuracy was then calculated using (23). Table V summarizes $\delta$, accuracy and training time for each test system. The MTLR and NN model provides high training and testing accuracy >95% for all the test systems considered in this work. Once the model is trained then the predictions, $u_{m,g,t}^{ML}$ and $P(u_{m,g,t}^{ML})$ for each test samples $m \in M^{test}$ are obtained and stored for all test systems.

### E. Verification Results (Online) & Feasibility Layer Benefits

In order to successfully assist SCUC, we developed the FL and the post-processing technique mentioned in sub-section IV.D and IV.E, respectively. The MTLR and NN based test predictions/outputs, $u_{m,g,t}^{ML}$ and $P(u_{m,g,t}^{ML})$ is verified for feasibility with FL to obtain $u_{m,g,t}^{MF}$ and then post-processed. To address model-reduction, benefits verification is performed for all test samples. The verification is an optimization step based on the ML outputs and therefore is an online step. Since the FL is also an optimization step, the solve time is inclusive of both post-processing and the MILP solve time. The R-SCUC-FL is implemented as per algorithm I in sub-section IV.E. In order to compare the benefits of FL, the R-SCUC (i.e without FL), is also performed. R-SCUC is implemented by replacing *step 2* and *step 3* in Algorithm I by:

- **Step II**: for remaining generators after Step 1, $fix\ u_{g,t}^m = 1$ if $P(u_{m,g,t}^{ML}) \geq 90\%$, $fix\ u_{g,t}^m = 0$ if $P(u_{m,g,t}^{ML}) \leq 10\%$ and warm-start $u_{g,t}^m = u_{m,g,t}^{ML}$ if $10\% < P(u_{m,g,t}^{ML}) < 90\%$.

Table VI represents the infeasible problems corrected with R-SCUC-FL by using MTLR and NN based ML outputs respectively. The infeasible problems arise in R-SCUC. Based on our study, we noted that R-SCUC resulted in infeasible problems in many samples since ML mainly cannot distinguish minimum up/down time physical constraint requirement for generators (8)-(9). It only requires incorrectly identifying one generator $g$ in time period $t$ to result in an infeasible solution for R-SCUC. For example, in IEEE 24-bus system, there are 33 generators and 24 time periods, which implies a total of 792 predictions per sample $m$ to identify commitment schedule. However, we notice that the FL eliminates all infeasible samples in all test systems. Here, NN R-SCUC is more susceptible to infeasible samples in R-SCUC in comparison to MTLR R-SCUC. But, irrespective of the ML model, the FL ensures that the ML outputs adhere to MILP requirements particularly the generator minimum on/off time limit constraints.

TABLE VI. FL INFEASIBLE PROBLEMS ELIMINATION

| System | IEEE 24-Bus | IEEE 73-Bus | IEEE 118-Bus | SG 500-Bus | Polish 2383-Bus |
|---|---|---|---|---|---|
| NN | 28 | 18 | 4 | 32 | 6 |
| MTLR | 4 | 6 | 0 | 8 | 4 |

Fig. 8 represents the solution quality whereas Fig. 9 represents the solve time averaged over all test samples for each test system utilizing the MTLR and NN based R-SCUC with and without FL. The objective cost and solve time for reduced models are represented as base-normalized (BN) values averaged over all test samples. The base model is normal SCUC that does not use any ML outputs. From Fig. 8, it can also be noted that the solution quality is maintained to a high degree for R-SCUC-FL without infeasibilities. In the case of IEEE 24-bus system, both MTLR R-SCUC and MTLR R-SCUC-FL result in better cost compared to SCUC. Similarly, the IEEE 73-bus system, NN R-SCUC and NN R-SCUC-FL result in lower cost. This is because model reduction on top of time-saving can also tighten the MIPGAP to a high degree resulting in a better MIPGAP solution in some test systems when compared to SCUC. However, on average across all test system, the solution quality is maintained to high degree of <0.1 % deviation for MTLR R-SCUC-FL and NN R-SCUC-FL.

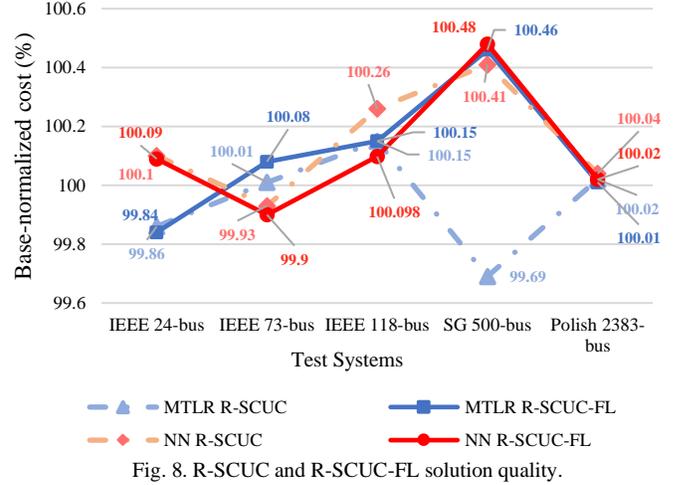

Fig. 8. R-SCUC and R-SCUC-FL solution quality.

The BN solve time shows that R-SCUC-FL requires a minor additional time for ML prediction post-processing as two MILP models are solved when compared with R-SCUC to ensure solution quality and eliminating infeasibility. MTLR R-SCUC-FL results in a speed-up factor of 1.7x, 3.3x, 2.1x, 2.3x and 8.5x, whereas NN R-SCUC-FL results in a speed-up factor of 1.6x, 3.7x, 1.9x, 2.8x and 6.9x for the IEEE 24-bus, IEEE 73-bus, IEEE 118-bus, SG 500-bus and Polish systems, respectively on average over all testing samples, $m \in M^{test}$, when compared with SCUC. When averaged across all test systems, MTLR R-SCUC-FL results in speed-up factor of 3.6x whereas NN R-SCUC-FL results in a speed-up factor of 3.4x while also ensuring feasibility of all test samples.

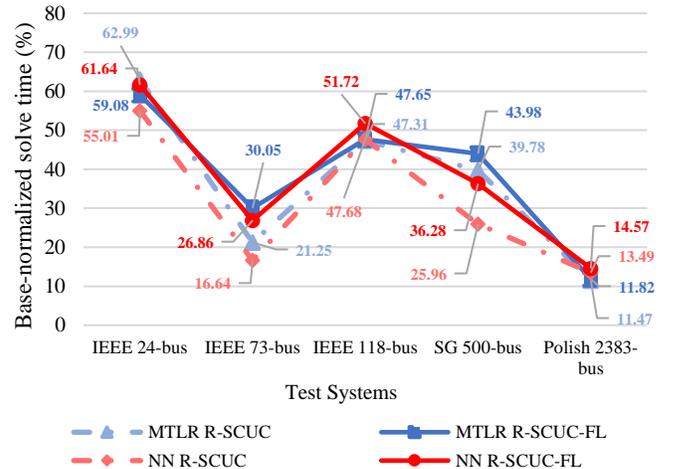

Fig. 9. R-SCUC and R-SCUC-FL solve time comparison.

### F. Out of Sample Testing

To understand the robustness of the proposed FL, an out-of-sample testing was further performed. The out-of-sample set consists of 100 samples that were not included in the training or testing samples of the verification process. Here, care was taken to introduce higher variability with $\alpha^m = \pm 25\%$ and $\beta_{n,t}^m = \pm 10\%$ in (16) in order to avoid mimicking the original dataset and increase number of infeasible samples.

TABLE VII. INFEASIBLE PROBLEMS IN OUT-OF-SAMPLE DATA

| Test System | MTLR R-SCUC | MTLR R-SCUC-FL | NN R-SCUC | NN R-SCUC-FL |
|---|---|---|---|---|
| IEEE 24-Bus | 40 | 14 (65% ↓) | 59 | 28 (53% ↓) |
| IEEE 73-Bus | 95 | 54 (41% ↓) | 100 | 74 (26% ↓) |
| IEEE 118-Bus | 63 | 27 (57% ↓) | 78 | 18 (77% ↓) |
| SG 500-Bus | 82 | 36 (56% ↓) | 100 | 67 (33% ↓) |
| Polish 2383-Bus | 37 | 9 (75% ↓) | 45 | 16 (64% ↓) |

These samples were never utilized in the offline training phase or online verification phase. Therefore, the trained model might not fare as well in the out-of-sample dataset (with much larger variations) when compared to the original dataset. Despite this, from Table VII, we notice a significant reduction in infeasible problems when the FL was introduced in R-SCUC in all test systems. This resulted in reductions of infeasible samples by 58.8% and 50.6% when averaged across all test systems for MTLR and NN models, respectively.

*G. Case Study: Multi-Scenario Renewable Source*

As stated in the prior section, the proposed MTLR methods are agnostic to the MILP model. Hence, it can be utilized for both stochastic-SCUC (SSCUC) and deterministic SCUC cases. In a deterministic scenario, renewable profile can be captured through net-load profile. However, the renewable energy is unpredictable in nature, the scenarios of wind and solar outputs are often considered. But it can be noted that in SSCUC, a single commitment schedule that satisfies all the scenarios are obtained as outputs. In terms of ML, this only increases the number of inputs but the targets/outputs remain the same. Therefore, the MTLR and NN models are modified to increase $S$ scenarios of net nodal load as input.

TABLE VIII. MODIFIED IEEE 24-BUS RENEWABLE SYSTEM RESULTS

| Metrics | MTLR R-SSCUC FL | NN R-SSCUC FL |
|---|---|---|
| Training Accuracy | 96.57% | 97.01% |
| Testing Accuracy | 94.53% | 96.25% |
| Infeasible samples | 0 | 0 |
| BN cost | 100.07% | 100.01% |
| BN Time | 43.57% | 36.78% |

The proposed MTLR R-SCUC FL and NN R-SCUC FL were tested on the modified IEEE 24-Bus renewable test case with two renewable units. Table VIII shows the online verification results. It can be noted that, the proposed MTLR and NN models can successfully handle stochastic inputs with solution accuracy of 94.53% and 96.25% for test samples. This is marginally lower than the deterministic case. However, utilizing the MTLR and NN solutions, we notice that the Reduced-SSCUC-FL (R-SSCUC) results in higher time savings of 56.43% and 63.22% with respect to SSCUC. In comparison, the deterministic MTLR and NN based R-SCUC-FL only results in a time saving of 40.92% and 38.36% with respect to SCUC. This is because, reducing equivalent number of variables benefits R-SSCUC more since this directly relaxes higher number of constraints when compared with R-SCUC.

VI. CONCLUSIONS

In this paper, we studied the differences between different classification techniques as an offline step namely, KNN, RF, LR and NN for predicting commitment schedules given the load profile. It was concluded that ML cannot directly replace optimization through SCUC. However, by choosing a confidence level through probabilistic outputs, selective binary variables were reduced in SCUC. LR and NN were more flexible due to the ability to result in probability estimates of commitment status of generators. Not only that, by studying the confusion matrix for ML predictions, both LR and NN led to higher accuracy and resulted in better predictions when compared to KNN and RF. Furthermore, LR was also addressed for computation efficiency through a novel MTLR model. On average, the MTLR model was 2.4x faster than LR during offline training.

The trained models were then introduced for online verification on test samples through post-processing ML solutions with FL. A confidence based variable selection and FL in combination produced desired effects of eliminating infeasible outputs while also maintaining high degree of solution-quality. On average across all test systems, model reductions with the proposed MTLR R-SCUC FL and NN R-SCUC-FL resulted in a speed-up 3.6x and 3.4x, respectively, when compared with SCUC.

On top of this, it was established that the proposed approach is agnostic of MILP models. Therefore, the ML model was also tested on a modified IEEE 24-bus system with three renewable scenarios. The ML outputs were then similarly used for variable reduction in SSCUC. Online verification of MTLR and NN based R-SSCUC-FL resulted in a speed-up of 2.3x and 2.7x, respectively, when compared to SSCUC. In comparison, the deterministic R-SCUC-FL resulted in a speed-up of 1.7x and 1.6x, respectively, when compared to SCUC for the IEEE 24-bus system. It is also worth noting that the proposed model reduction approaches are compatible with any existing optimization/decomposition methods as well as ML methods aiming to remove some unnecessary constraints.